**Long-time temperature variations in Portugal over the last 140 years and the effect of the solar activity**

Anna L. Morozova [a] [1]*
[a] Astronomical Observatory of the University of Coimbra, Santa Clara, 3040-004 Coimbra, Portugal (anita_geo@hotmail.com)

Peter Thejll [b]
[b] Danish Climate Centre at the Danish Meteorological Institute, Lyngbyvej 100, DK-2100 Copenhagen, Denmark (pth@dmi.dk)

Maria Alexandra Pais [c]
[c] CFC, Department of Physics, University of Coimbra, 3004-516 Coimbra, Portugal (pais@fis.uc.pt)

* Corresponding author
[1] Visiting Astronomer

**Abstract**

We present the analysis of temperature variations in Portugal for 140 years (from 1865 to 2005). The two stations with the longest time series of temperature measurements (Lisbon and Coimbra) have been used to study the dependence of the portuguese climate variations on the changes of both global circulation and solar activity. Monthly averaged temperature series have been analyzed together with monthly North-Atlantic Oscillation (NAO) index data, sunspot numbers (W) and cosmic ray (CR) flux intensity. Different statistical methods (the correlation analysis and the multiple regression analysis) were used. Our results show that the temperature in Portugal depends not only on the atmospheric variations in the studied region but also on the variations of the solar parameters such as sunspot numbers and the cosmic rays flux intensity. Also, the dependence of temperature on solar parameters is strong during the cold season (November to February), while much weaker during the warm season. For some months, strong correlations between the temperature variations of the current month and the North-Atlantic Oscillation index (NAOI) values of the previous month have been found. The correlation between climatic and solar parameters shows up best on the decadal and decadal-to-centennial timescale. It is found that the temperature correlates positively with the sunspot numbers and negatively with the CR flux intensity throughout the year. Besides, the absolute values of the correlation coefficients between the temperature and the CR are higher than those between the temperature and the sunspot numbers. Our results are consistent with some of the proposed mechanisms that relate solar activity to Earth climate and could be explained through the effect of the solar UV radiation and stratosphere-troposphere coupling or/and through the effect of the CR particles on clouds and stratospheric and tropospheric conditions.

**1. Introduction**

Climate is most certainly affected by several natural (such as atmospheric composition, solar energy flux, albedo), and anthropogenic (e.g. atmospheric pollution) factors. However, there are also subtle natural phenomena which can have a significant impact on the climate. One of such phenomena is solar activity. The pioneering papers (e.g. Lawrence, 1965; Ney, 1959) concerning the effect of solar activity on climate were published in the sixties. Later, a significant number of studies have been performed in that field reporting correlations between parameters characterizing solar activity and numerous meteorological parameters such as temperature, pressure, wind velocity and global circulation, cloudiness and atmospheric transparency and precipitation (see reviews by Carslaw et al., 2002; Haigh, 2003; Rind, D. 2002; Tsiropoula, 2003 and references therein). These correlations have been studied on long-time scales (e.g. the 11-year and 22-year solar cycles, see Blackford and Chambers, 1995; Svensmark and Friis-Christensen, 1997; Thejll et al, 2003; Walsh and Kleeman, 1997) as well as on short-time scales when the solar activity variations last some days (Kniveton, 2004; Pudovkin et al, 1996, 1997; Veretenenko and Thejll, 2004).

There are three main mechanisms describing the possible influence of the solar activity on the Earth's weather and climate. These mechanisms are associated with
1) *Variations of the total solar irradiance (TSI).* These variations are in strong correlation with the sunspot number variations, but the amplitude of these variations is very small (~0.1% of the TSI mean value during the 11-years solar cycle – Fröhlich, 2000, 2003, 2004; Fröhlich et al., 2006).

2) *Variations of the UV part of the solar spectrum and UV absorption by the ozone in the stratosphere*. The variations of solar UV irradiance are also in strong correlation with the sunspot number variations and have significant relative amplitude (from 3% to 20% for different wavelengths during the 11-years solar cycle). These variations affect first the stratosphere (due to ozone absorption) and then influence the tropospheric conditions through the stratosphere-troposphere coupling (Baldwin and Dunkerton, 2005).

3) *Variations of the CR flux intensity*. CR flux is directly controlled by the solar activity; its variations are in strong anti-correlation with the sunspot number variations and have significant relative amplitude (tens of percent during the 11-years solar cycle and about 11% during the last 150 years). The CR particles are one of the main sources of ionization in the middle to low atmosphere. They are expected to affect the electrical properties of the atmosphere with possible implications for cloud droplets nucleation causing an increase in cloudiness in the troposphere and stratosphere thus changing the heat distribution in the lower atmosphere. A more thorough description of these mechanisms can be found in e.g. Carslaw et al., 2002; Haigh, 2003; Marsh and Svensmark, 2000, 2003; Rind, 2002; Svensmark and Friis-Christensen, 1997; Svensmark et al., 2009; Tinsley and Deen, 1991; Pudovkin et al, 1996, 1997, though some details in this process have to be studied thoroughly (Kristjánsson et al., 2008; Sloan and Wolfendale, 2008).

There is another possibility for the CR particles to affect the Earth's troposphere which is related with the above mentioned mechanism #2: through the modification of the stratospheric ozone level. The CR both produce and destroy ozone and there is a certain balance between these two opposite processes which seems to depend on latitude (Krivolutsky et al., 1999). Thus, long-term variations of ozone amount related to CR variations are clearly possible (see e.g. Krivolutsky, 2003) for the review on the interaction between CR and the Earth's atmosphere). Since the temperature of the stratosphere depends on its constituent gases, the periodic changes of the CR flux may change the stratospheric temperature and affect stratospheric circulation (Hood, 2004). Such changes of the stratospheric conditions could affect the upward propagation of planetary-scale waves, and lead to an indirect feedback on the lower atmosphere (Kodera and Kuroda, 2002). The mechanisms of energy transfer from the upper atmosphere to the troposphere are being studied recently (Baldwin and Dunkerton, 2005; Haigh, 2009; Lean, 2010).

Presently, one of the main issues in studies of the effect of solar activity on the Earth's weather and climate is the latitudinal and longitudinal dependence of the variations of the atmospheric parameters on solar activity (e.g. Le Mouël et al., 2009). One of the main factors which cause this dependence is the latitudinal variations in the Earth's magnetic field and, consequently, in the intensity of the CR flux in a certain region of the Earth. Another important factor that contributes to the geographical variations of atmospheric parameters is the effect of predominant local climatic conditions (proximity to the sea or mountain ranges, air pollution and aerosol opacity and the patterns of the atmospheric circulation – see e.g. Voiculescu et al. (2006). Several studies have already been performed to evaluate the difference in the response of the troposphere to the solar activity forcing for different latitudinal and longitudinal regions (see e.g. Dobrica, 2009; Gleisner and Thejll, 2003; Kristjánsson et al, 2008; Le Mouël et al., 2009; Morozova et al., 2002; Voiculescu et al., 2006). This type of studies shows the geographical patterns of the tropospheric response to the solar activity variations and brings further insight into the physical mechanisms which are involved.

Another climatic feature which is currently being studied is the seasonal dependence of the atmospheric response to solar activity forcing. It has been shown in several studies that during winter in the Northern Hemisphere the atmosphere is dynamically more active than during summer, especially in Europe, due to the strong influence of the NAO from November to March (see e.g. Hurrel et al., 2003; Visbeck et al., 2001), and strongly responds to the variations in the solar parameters (as reported in e.g. Le Mouël et al., 2009).

In this paper we present an analysis of the variations of temperature in Portugal during a 140 years interval (from 1865 to 2005) and relations of the climate in this region to the variations of the solar activity. We have used several statistical methods to study the level and the significance of these relations. The climatic and solar parameters, data processing and statistical methods are presented in Sections 2 and 3. Sections 4 and 5 are dedicated to the presentation and discussion of the obtained results. Physical mechanisms which could explain the observed relations between the climatic and solar parameters are discussed in Section 6.

## 2. Data
### 2.1. Data list

The local climate was analyzed using the following data:
1) monthly means of temperature (T) recorded by two portuguese meteorological sites: one near Lisbon (observation period from 1855 to 2008, kindly provided by Dr. M. A. Valente (IGIDL)) and the other near Coimbra (observation period from 1865 to 2005, kindly provided by Dr. I. Alves and M. J. Chorro (IGUC)).
2) The NAOI has been used to characterize the atmospheric circulation in this region. The NAOI used in this analysis is defined by Li and Wang (2003) as the differences of normalized sea level pressures regionally zonal-averaged over a broad range of longitudes 80W–30E (see Li and Wang, 2003 for full description).The data set contains monthly data from 1873 to 2005 (http://www.lasg.ac.cn/staff/ljp/data-NAM-SAM-NAO/NAO.htm).

The variations of the solar activity were analyzed using the following data:
1) monthly averaged sunspot numbers (W) from 1855 to 2005 taken from NGDC database (ftp://ftp.ngdc.noaa.gov/STP/SOLAR_DATA/SUNSPOT_NUMBERS);
2) monthly averaged data set on the cosmic rays flux from Climax neutron monitor (NM) from 1953 to 2005 taken from NGDC data base (ftp://ftp.ngdc.noaa.gov/STP/SOLAR_DATA/COSMIC_RAYS/STATION_DATA/Climax)
3) reconstructed annual data of the CR variation from 1865 to 2005 (see Alanko-Huotari et al., 2006; Usoskin et al., 2002, 2005, 2008 for full description) taken from NOAA data base ftp://ftp.ncdc.noaa.gov/pub/data/paleo/climate_forcing/solar_variability/usoskin2008crii.txt.

## 2.2. Climatic Data

The climate in Portugal is strongly influenced by the circulation modes of the atmosphere over the North Atlantic region. The strength and the direction of the westerly winds and storm tracks across the North Atlantic depend on the sea-level pressure difference between the Icelandic Low and the Azores High (Xoplaki, 2002). The effect of the NAO on the European climate is stronger during the months from November to March/April when the atmosphere is dynamically most active (Hurrel et al., 2003; Slonosky and Yiou, 2002; Visbeck et al., 2001). During other seasons the correlations between the sea level pressure (as well as the surface air temperature) and NAOI are much weaker, showing that during those seasons the NAOI can not *fully* characterize the atmospheric circulations in the North Atlantic region (Li and Wang, 2003).

Previously, some authors (see e.g. Miranda et al., 2002 and references therein) reported the correlations between the NAOI and the mean pressure and precipitation in Portugal. Also, a weak connection between the NAOI and the temperature has been found in many other studies (see e.g. Luterbacher and Xoplaki, 2003, Sáenz et al., 2001a, 2001b). When NAO is in its negative phase, the Iberian Peninsula is under the strong influence of the Atlantic cyclones. The cyclones bring humid air increasing the precipitation amount in this region. However, the temperature of these air masses depends on the current wind direction (Prieto et al., 2002). Besides, there is also a dependence of the temperature in Portugal on the East Atlantic (EA) mode of the atmospheric circulation as shown by Sáenz et al. (2001a, 2001b). Thereby, the temperature variations depend both on the NAO (Sáenz et al., 2001a, 2001b) and on the circulation patterns that are created as a result of the interaction of the EA and the NAO modes (Prieto et al., 2002; Trigo et al., 1999). In this study we use the NAOI as a rough proxy of the atmospheric circulation mode over Portugal.

## 2.3. Solar Data

Different solar parameters - sunspot number, total solar irradiance (TSI), spectral solar irradiance at radio, UV and X-ray wavelength - correlate strongly with each other on the yearly to decadal time scales. The TSI variations during the 11-year solar cycle have amplitude of about 0.1% and are in phase with solar activity variations. The UV part of the solar spectrum shows higher levels of variability: during the 11-year solar cycle the solar irradiance at wavelength of 140, 200 and 250 nm changes in phase with solar activity with amplitude of 20%, 8% and 3%, respectively (see Fröhlich, 2000, 2003, 2004; Fröhlich and Lean, 1998a, 1998b; Lean, 1997; Tsiropoula, 2003). The time periods during which the direct measurements of the TSI and UV irradiance were conducted (about 30 years, Fröhlich, 2004) do not allow to analyze connections between the temperature and the solar irradiance variations on the decadal and centennial time scales. However, the strong correlation between the sunspot numbers and both irradiance indices during the above mentioned periods allows us to use the sunspot numbers as a proxy for irradiance variations over decadal and centennial time scales.

The longest time series of the CR is provided by the ground-based Climax neutron monitor in Colorado, USA. The CR data show strong anti-correlation with the sunspot numbers (see Table 1), yet these two parameters are not interchangeable. There is a difference in phase and amplitude of the 11-year variations of the sunspot numbers and CR flux and this difference varies over time. On the centennial timescale this difference is even clearer. The Climax data set starts in 1953. However, a longer time series is needed to study the long time scale relations between the variations of the meteorological and solar parameters. Therefore, for the further analysis a reconstructed CR data series has been used. The reconstruction of the CR flux is based on a model that includes sunspot series, calculation of the solar magnetic flux and heliospheric modulation of the CR flux, and a comparison to measured $^{10}$Be data (see Alanko-Huotari et al., 2006; Usoskin et al., 2002, 2005, 2008 for full description).

## 3. Data preparation and statistical methods

Due to the proximity of Coimbra and Lisbon (~ 200 km) the meteorological data from these two sites show a strong correlation. In order to combine two very similar temperature records from two nearby stations we performed Empirical Orthogonal Function (EOF) analysis to extract the leading principal components, and used this in the following analysis. EOF analysis (or Principal Component Analysis - PCA) is a mathematical procedure that transforms a number of possibly correlated variables into a smaller number of uncorrelated variables called principal components (PCs). The first principal component accounts for as much of the variability in the data as possible, while each succeeding component accounts for as much of the remaining variability as possible and the PCs are all independent to each other. Also, the EOF analysis helps to improve signal to noise ratio: only the PCs with sufficiently high eigenvalues correspond to real signal, common to both series.

The two Principal Components (PC1 and PC2, normalized) of the temperature data from the two portuguese sites were calculated. Figure 1 shows the original annual data series together with the PC1 (Fig. 1a) and the PC2 (Fig. 1b) variations. The first principal component corresponds to the largest eigenvalue. The PC1 (Fig. 1a) has significant amplitude and for different months accounts for 93-97.5% of the original temperature variations observed at both stations. The PC2 (Fig. 1b) has small amplitude, it shows no significant correlations with the original temperature data (correlation coefficient $|r| \leq 0.26$) and in terms of this study it could be considered as noise. Therefore, the PC1 was used in the following analysis instead of the two original temperature data sets.

The climatic data (T and the NAOI) for different months were analyzed separately. This allows us to exclude the seasonal cycle from the data and to study the variations of the climate for different months and seasons separately. To study the relations between the climatic and the solar parameters, non-smoothed as well as smoothed data were used. The data were smoothed by the 5 and 11 years adjacent averaging procedure: the smoothed value at index *i* is the average of the data points in the interval from *i-2* to *i+2* or from *i-5* to *i+5*, respectively . The smoothing procedure allowed us to emphasize the decadal and centennial variations of the solar and atmospheric parameters. Both non-smoothed and smoothed data were analyzed using correlation analysis and multiple regression analysis.

To study decadal variations in the climatic parameters, which are possible imprints of the 11-year solar cycle, the non-smoothed and 5 years adjacent averaged climatic data were correlated with the non-smoothed solar data: in this case both solar and climatic data show the decadal variations. To study the relations between the decadal-to-centennial variations and the trends in the Earth's climate and in the solar activity, the climatic data smoothed by the 11 years adjacent averaging were correlated with the 11 years adjacent averaged solar data: in this case both solar and climatic data show the decadal-to-centennial variations.

To describe the significance levels of our statistical results we use the *p*-level value which is probability to observe such correlation coefficient due to chance. In this case the lower is p-level value the more significant is corresponding correlation coefficient. The smoothing procedure causes an increase of the autocorrelation of the data –neighbouring data points become more correlated. Therefore, if the smoothed data are subjected to statistical analysis, the significance of the statistical parameters has to be calculated using "effective sample sizes" instead of the actual sample sizes (Hsieh, 2008). The effective sample size value gives a number of the statistically independent points in the analyzed data sets and for each pair of the analyzed parameters it was calculated as

$$n_{eff} = N/(1 + 2\sum_{i=1}^{N} a_i a'_i),$$

where $n_{eff}$ is the effective sample size, $N$ is the actual sample size (141 years for the temperature and solar data and 133 years for the NAOI), $a_i$ and $a'_i$ are the values of the autocorrelation functions of the analyzed data sets with lag $i$ (Hsieh, 2008; Smith et al., 1996). For temperature (NAOI) smoothed by 5 and 11 years adjacent averaging the calculated $n_{eff}$ are in the range 19-73 (25-58) and 4-39 (7-18) years, correspondingly. The use of the effective sample size, which is always smaller then the actual sample size, reduces the significances of the calculated correlation coefficients, but, which is more important, helps to avoid spuriously high significances which would have appeared without reducing the number of degrees of freedom due to smoothing procedures.

The calculation of the correlation coefficients significances using $n_{eff}$ shows low significance of the correlation between smoothed data. However, as any of parametric statistical methods, this method of the significance calculation relies on assumptions that the data are drawn from a given probability distribution, and these assumptions are not necessary correct for a real climatic or solar data series.

There are others, non-parametric, methods to calculate the significance of calculated correlation coefficients. One of them is Monte Carlo simulations on surrogate data generated (in our case) from solar as well as climatic data (Theiler and Prichard, 1996; Thejll et al., 2003). This method allows one to keep the autocorrelative structure of the original series. The method could be described briefly as follows. Let $X_i$ and $Y_i$ be two data series with correlation coefficients $r_0$. We can construct artificial (or surrogate) data series $X'_i$ (or $Y'_i$) that have the same statistics (mean values, standard deviations and auto-correlations) as the real data and calculate the correlation coefficient $r'$ between $X'_i$ and $Y_i$ (or $X_i$ and $Y'_i$). By applying this procedure $N_{MC}$ times, we obtain a series of $r'_n$ coefficients. The significance of the correlation coefficients $r_0$ between two original series $X_i$ and $Y_i$ is then calculated as

$$p(X',Y) = n_r/N_{MC},$$

where $n_r$ is a number of realizations with $|r'_n| \geq |r_0|$ (for double sided hypothesis testing). In our analysis $N_{MC} = 10,000$ and $p = (p(X',Y) + p(X,Y'))/2$.

For generating surrogate data, we fitted an autoregressive moving average (ARMA) model to the original data. The ARMA models are often used to describe the variability of climatic and solar data due to the presence of autocorrelation and white noise (Bershadskii, 2009; Gil-Alana, L. A. 2009; von Storch, H., 1995). For a given data series $X_i$ the ARMA (f, q) model is written as

$$X_i = c + \sum_{k=1}^{f} \varphi_k \cdot X_{i-k} + \sum_{k=1}^{q} \theta_k \cdot \varepsilon_{i-k} + \varepsilon_i$$

where $f$ is the order of the autoregressive part, $q$ is the order of the moving average part, $c$ is a constant, $\varphi_k$ and $\theta_k$ are autoregressive and moving average parameters, respectively, and $\varepsilon_k$ are white noise error terms. In our analysis the surrogate data are created by the ARMA (2,2) model with coefficients $\varphi_k$ and $\theta_k$ taken from the corresponding real data sets (based on the maximum likelihood estimation method).

The comparison of the significances calculated by "effective sample size" and Monte Carlo methods shows that for non-smoothed data both methods give almost identical results, for 5 years adjacent averaged data Monte Carlo method gives similar or slightly higher significance levels and for 11 years adjacent averaged data the Monte Carlo method gives lower significances for small correlation coefficients and higher significances for large correlation coefficients. The significances of all correlation coefficients shown in this paper (Figures 2 and 3; Tables 2 and 3) are calculated using the Monte Carlo method and are referred thereafter as "*individual significances*".

The other problem that arises during the estimation of the significance of the obtained correlation coefficients is multiple comparisons. It occurs when the same statistical calculations are applied to a number of data sets simultaneously: the larger the number of simultaneously analyzed data sets, the easier to obtain by chance a result with high individual significance (Abdi, 2007). Therefore, a special correction of the significance level of the obtained statistical parameters is needed.

In our case, as a rule, we estimate the correlation between a series of solar data and 12 series of monthly climatic data, obtaining a set of 12 correlation coefficients. Another series of Monte Carlo simulations was performed to calculate the corrected significance (referred thereafter as "*batch significance*") of the obtained correlation coefficients. Each time we generate 12 surrogate series $X''_{1i},…, X''_{12i}$ corresponding to 12 series of monthly climatic data (or one series $Y''_i$ corresponding to a series of solar data) and calculate a set of 12 correlation coefficients $r''_1, …, r''_{12}$ between $X''_{1i},…, X''_{12i}$ and $Y_i$ (or $X_{1i},…, X_{12i}$ and $Y''_i$ ) and choose the correlation coefficient with the largest absolute value $r''_{max}$. By repeating this procedure $N_{MC}$ times, we obtain a series of $r''_{max\,n}$ coefficients. The significance of any correlation coefficient $r_{0\,m}$ from a set of $r_{0\,1}, …, r_{0\,12}$ is calculated as
$$p_m(X'',Y) = n_m/N_{MC},$$
where $n_m$ is a number of realizations with $|r''_{max\,n}| \geq |r_{0\,m}|$. In our analysis $N_{MC}$ = 10,000 and
$p = \bigl(p(X'',Y) + p(X,Y'')\bigr)/2$.

As expected, for most correlation coefficients batch significances are much lower then individual significances (bigger *p* values). However, in some cases the batch significances are high with $p \leq 0.2$ or even $p \leq 0.1$ (see Figs. 2a-b and Figs. 3a-d and Sections 4 and 5).

**4. Results for dependence of the temperature on the atmospheric circulation.**

Fig. 2a shows the correlation coefficients between the temperature in Portugal and the NAOI for different months of the year. The data show significant correlations for some months, though, the sign and the values of the correlation coefficients vary with the season. As one can see, the non-smoothed data show a weak correlation: $|r| \leq 0.27$ with the highest individual significances (lowest *p*-levels). The smoothed data show stronger correlation values: for data smoothed by the 5 years adjacent averaging $r \leq 0.47$ with individual p-levels from 0.03 to 0.26 and for data smoothed by the 11 years adjacent averaging $r \leq 0.55$ with individual p-level from 0.06 to 0.46. During some months the PC1 and the NAOI show anti-correlation (January and November) or no correlation at all (mainly, in spring-summer). Also, there is a strong significant positive correlation between the temperature in Portugal and the variations in the NAOI on the decadal and centennial time scales and only for specific months (February, April, August). Numbers in Fig.2a show the batch significances (see Section 3) of the largest correlation coefficients for non-smoothed and smoothed data: the probability to get by chance such large correlation coefficient in a set of 12 series of monthly data is 0.03 for non-smoothed data and 0.11 for 5-years adjacent averaged data.

Fig. 2a illustrates in particular the relations between the temperature (PC1) and NAOI during the four cold months: from November to February. As one can see, during January and November there is a weak anti-correlation between the air temperature and the NAOI. However, during February these parameters correlate and during December no significant correlation was found. As shown below (Section 5.2), the correlation analysis of the solar and atmospheric parameters allowed us to distinguish a "cold" (November to February) and a "warm" (June to September) seasons during which significant correlations between the solar and the atmospheric parameters have been established. However, since as we can see in Fig. 2 the temperature dependence on the NAOI during this "cold" season changes from one month to another, the correlation between the PC1 and the NAOI for the whole "cold" season is close to zero (see Table 2).

We have also found that for some months the correlations between the temperature during the month under consideration and the NAOI of the *previous month* are higher than the correlation between the temperature and the NAOI of the same month. Figure 2b shows the corresponding correlation coefficients and their individual and batch significances. The strongest and most significant correlation is obtained for the pair $T_{January}$ vs $NAOI_{December}$: $r$ is from -0.26 to -0.7, individual significance is from 0.001 to 0.007 and batch significance is from 0.01 to 0.04. Similar dependence has been described by the Sánchez et al. (2007): they showed that the NAOI variations lead sea surface temperature anomalies near the Iberian Peninsula by 1 month. This feature of the T-NAOI relations is taken into account below in the model of the multiple regressions (Section 5.3).

## 5. Results for relations between the atmospheric and solar parameters
## 5.1. Correlation analysis (all months)

As a first step in the analysis of the relations between the atmospheric (the temperature and the NAOI) and the solar (the CR flux and the sunspot number W) parameters, we calculated the correlation coefficients for non-smoothed and smoothed (by 5 and 11 years adjacent averaging) data. The correlation coefficients between the climatic and solar parameters (and their significances) are given in Figs. 3a-d. The results of the correlation analysis show that there are similarities between the variations of the climatic and the solar parameters. These similarities are manifested most clearly on the decadal timescale (smoothed data) and almost disappear on the year-to-year timescale. One of the possible explanations of this behavior is a year-to-year internal climate variability that conceals the long-term effect. Also, Figs 3a-d show that on the decadal time scale (e.g. 11 year cycle) the relations between the climatic and solar parameters are mainly weak or non-significant. On the contrary, the data smoothed by the 11 years adjacent averaging show that on the decadal-to-centennial time scale there are strong significant relations between the climatic and solar parameters. Numbers in Fig.3a-d show the batch significances of the largest correlation coefficients for non-smoothed and smoothed data. In the case of correlation between temperature and solar parameters, the probability to get by chance such large correlation coefficient in a set of 12 series of monthly temperature data is from 0.09 to 0.20 for non-smoothed data, 0.18-0.22 for 5-years adjacent averaged data and 0.17-0.34 for 11-years adjacent averaged data. In the case of correlation between NAOI and solar parameters, the same probability for monthly NAOI data is very high: 0.82 for 5-years adjacent averaged data and 0.66-0.76 for 11-years adjacent averaged data and means that in this particular case our results could not be treated as statistically significant.

The relations between the atmospheric and solar parameters are seen most clearly during the cold months (November to February) and become weaker in the rest of the year. There is a positive correlation between the temperature and the sunspot numbers variations and a negative correlation between the temperature and the cosmic rays flux variations. Correlation analysis shows that relations between the solar parameters variations and the temperature are stronger and more significant (Figs. 3b and 3d) than relations between the solar parameters and the NAOI (Figs. 3a and 3c). The correlation coefficients are not very high. That means that the solar activity is not a determining factor for the climate variations but can be a modulator and/or trigger (see discussion in Carslaw et al., 2002; Haigh, 2003; Rind, D. 2002;

Tsiropoula, 2003 and references therein). The other possible explanation for the small correlation coefficients is that the internal atmospheric processes damp or mask the solar effect. Still, the data shown in Figs. 3a-d allow one to explain (in a statistical sense) up to 44% of the decadal-to-centennial variance of the temperature during some months by the changes of the solar activity. The interesting feature is that the 11-year solar cycle shows itself in the temperature variations in Portugal less than it does in more northern parts of the Europe (see Section 3.2 in Tsiropoula (2003) for a detailed description).

### 5.2. Correlation analysis (seasons)

The correlation analysis for the monthly data (see Figs. 3a-d) allows us to distinguish between a "cold" (November to February) and, a "warm" (June to September) seasons during which there are significant correlations between the solar and the atmospheric parameters. This effect is not totally unexpected: the stronger response of the winter atmosphere to the solar influence has already been reported (see e.g. Le Mouël et al., 2009). Table 2 shows the correlation coefficients calculated for the cold and the warm seasons using both non-smoothed and smoothed data. The individual significance of the correlation coefficients was calculated using Monte Carlo method as described in Section 3. The analysis of the data shows that during the cold season there is a significant correlation between the decadal and decadal-to-centennial variations (smoothed data) of the temperature and the solar parameters: $|r| = 0.2 - 0.5$ with $p$-levels from <0.001 up to 0.2 for 5 years adjacent averaged data and $|r| = 0.28 - 0.7$ with p-levels from 0.02 up to 0.46 for 11 years adjacent averaged data. On the whole, there is a positive correlation between the temperature and the sunspot numbers variations and a negative correlation between the temperature and the cosmic rays flux variations.

NAOI shows weak correlation with both solar parameters even on decadal-to-centennial timescale: for 11 years adjacent averaged data $r_{W\ vs\ NAOI} = -0.53$ with individual significance 0.14 and $r_{CR\ vs\ NAOI} = 0.4$ with individual significance 0.13.

### 5.3. Multiple regression analysis

As the next step, the multiple regression analysis was used to study the *individual* contribution of the NAOI and the solar parameters to the temperature variations. As independent parameters the NAOI of the current month ($NAOI_m$), the NAOI of the previous month ($NAOI_{m-1}$), the sunspot numbers (W) and the reconstructed cosmic ray intensity (CR) were used. The standard statistical parameter $R^2_{adjusted}$ was calculated. The ($R^2_{adjusted} \times 100$) values show the percent of the variability of the dependent variable (the temperature) that has been accounted for by the model under consideration. The set of the parameters used for each particular regression procedure was determined by the "best subset" technique based on the F-test that finds subsets of predictor variables that best predict responses on a dependent variable by linear regression. The $R^2_{adjusted}$ value was used as the criterion for choosing the best subset of predictor parameters. The particular sets of the independent parameters used for each regression procedure are shown in Table 3. As one can see, of all the independent parameters used in our analysis, the CR and $NAOI_m$ are the most influential (they are most often included in the "best subset"). As it is shown in Table 3, for most months there is a dependence of the decadal temperature variations on the atmospheric circulation in the North-Atlantic region described by the NAOI of the current and/or previous months. Also, there are strong relations between the temperature and the cosmic ray intensity for most months (except May and November).

The same analysis has been performed for the cold and the warm seasons as determined above. As independent regressors the NAOI of the current season ($NAOI_m$), the sunspot numbers (W) and the reconstructed cosmic ray intensity (CR) were used. On the decadal timescale during the cold season the temperature depends mainly on the cosmic rays, but during the warm season the temperature depends on all three independent regressors (see Table 3). Fig. 4 and Table 4 show the ($R^2_{adjusted}$ × 100) values calculated for different months (Fig. 4) and seasons (Table 4). As one can see, the multiple regression models based on the combination of the NAOI, CR and W variations can explain up to 72% of the temperature variance during individual months or up to 60% of the temperature variance during individual seasons. Again, the decadal and decadal-to-centennial temperature variations show stronger dependence on the NAOI and solar parameters (CR and W).

## 6. Discussion

The length of the used data sets (133 for the NAOI and 141 years for the temperature) does not allow us to estimate the statistical significance of the our results by the standard statistical methods (von Storch, H., 1995) e.g. by dividing the full data set into two subsets: one for the construction of the hypothesis and one for the testing. In our case, the division of the full data set into two parts would make it impossible to study the decadal and decadal-to-centennial variations. Nonetheless, we can use other methods to estimate the reliability of our results. First of all, we use Monte Carlo simulations to calculate both individual and batch significances of the obtained correlations. Secondly, our results are in agreement with known atmospheric features in the region under consideration: the weak dependence of the temperature in Portugal on the NAO mode of the atmospheric circulation during warm season becomes stronger during cold season as was reported in Luterbacher and Xoplaki (2003), Miranda et al. (2002), Sáenz et al. (2001a, 2001b). Finally, the results of applying two different methods (the correlation analysis and the multiple regression analysis) agree with each other. They both show the strong influence of the CR and NAOI on the variations of the temperature. They both show stronger dependence of the temperature on the solar parameters during cold season. They both show that the dependence of temperature on the NAOI and solar parameters is stronger on the decadal and decadal-to-centennial time scale than on inter-annual timescale.

In Section 1 we described three main mechanisms of the solar activity influence on the Earth's climate. The results of our study could be interpreted within these models:

1) *Variations of the total solar irradiance (TSI)*. As we found, temperature in Portugal correlates with sunspot numbers. Since TSI also correlates with sunspot numbers, it is possible to assume the following scheme: the higher the solar activity level, the more solar irradiation, the higher the temperature. However, the observed amplitude of the TSI variations (ca. 0.1%, Haigh, 2003) is sufficient only to account for a global temperature change of about 0.1 ºC (Rind, 2002) and is not enough to explain the temperature variations $\Delta T \approx 2$ ºC and corresponding $\Delta PC1 \approx 4.25$ (normalized) observed in Portugal. We believe that in the absence of a known significant internal atmospheric effect that could amplify the solar irradiance effect on tropospheric temperature, this mechanism could not be considered as sufficient to explain the temperature variations observed in Portugal.

2) *Stratospheric absorption of the solar UV radiation by the ozone and stratosphere-troposphere coupling*. Since the UV solar irradiance variations correlate with sunspot numbers, the following scheme is possible: the higher UV flux, the higher the stratospheric temperature and the ozone level, the stronger the heating of the troposphere by the stratosphere. However, since there is a strong anti-correlation between the temperature in Portugal and the CR flux variations, it also makes possible the following scheme: the less CR particles, the higher the ozone level (because CR particles can lead to depletion of the ozone layer), and the stronger the heating of the troposphere by the stratosphere. Accordingly, the stratospheric mechanism could explain the observed relations between the temperature and the solar parameters.

To evaluate the implication of a mechanism involving the stratosphere, we performed a dedicated study to analyze the effect of the quasi-biennial oscillations (QBO) of the equatorial zonal wind in the stratosphere. The importance of the phase of the QBO on the stratospheric response to the solar forcing has been shown in the studies by Labitzke and van Loon (e.g. Labitzke, 1987, 2005; Labitzke and van Loon, 1988). They found that during the west phase of the QBO, stratospheric temperature correlates with sunspot numbers, and during the east phase of the QBO the temperature and solar parameters tend to anti-correlate or non-correlate. To test the hypothesis about involvement of the stratospheric mechanism, we used data on the zonal wind at level 30 hPa (a combination of data from three radiosonde stations Canton Island, Gan/Maledives and Singapore, see Naujokat (1986) for details) from http://jisao.washington.edu/data/qbo/index.my_page.html. We divided our monthly PC1 data into two groups depending on the phase of the QBO (west or east) during the current month and year. After this, both groups were subjected to the correlation analysis as described above. The results (not shown in this paper) show no dependence on the QBO phase. However, this negative result is not sufficient to discard the influence of the stratosphere-troposphere coupling because the stratosphere-troposphere system shows strong internal variability and now it is not possible to make an accurate prediction of the influence of one parameter (e.g. QBO phase) on another (e.g. tropospheric temperature) – see e.g. Holton (1995).

3) *Variations of the CR flux intensity and cloudiness.* We found that there is a strong anti-correlation between the temperature in Portugal during the winter season and the CR flux variations. We could envisage a mechanism whereby the CR flux involves
 in the following way: the increasing CR flux, through the effect of some of the above discussed mechanisms, causes the intensification of the zonal circulation in the North Atlantic region. During the winter season the intensification of the zonal circulation (the increasing of the NAOI) means that the weather in Portugal would be cold and dry instead of mild and wet (Visbeck et al., 2001). That is in agreement with the observed data: the negative correlation between the NAOI and the temperature was found for November and January (Fig.3). This model also predicts the opposite effect during the summer season: increase of the CR flux and the intensification of the zonal circulation should result in hot and dry summers in the Iberian Peninsula. However, in our study the positive correlation between the NAOI and the temperature was found only for transitional periods (April-May and August-September). The probable explanation for this fact is that the temperature variations in Portugal show no significant correlation with the NAOI during summer season (Hurrel et al., 2003; Slonosky and Yiou, 2002; Visbeck et al., 2001; Xoplaki, 2002). It is also possible that dependence of the temperature in Portugal on CR variations could be explained by some other model that does not include strong effect of the NAO: e.g. through the influence of the EA mode of atmospheric circulation or the effect of the clouds formed directly in the region.

Thus, the results of this study could be explained using one (or both) of the following mechanisms of the solar activity influence on the Earth's weather and climate (mechanisms #2 and #3): 1. through the effect of the UV radiation on the stratosphere and stratosphere-troposphere coupling; 2. through the effect of the CR particles on the cloudiness and stratospheric and tropospheric conditions. Additional studies are necessary to conclude which of the mechanisms is more influential in the region under consideration. For example, the data presented in Voiculescu et al., 2006 show that in the western region of the Iberian Peninsula a low-level cloud amount (LCA) anti-correlates with the UV flux variations, but does not depend on CR variations.

As was mentioned above, the imprint of the 11-year solar cycle on decadal temperature variations in Portugal founded in this study is not so significant as in more northern parts of the Europe or North

America (Tsiropoula, 2003). On the other hand, computer simulations of the effect of the solar activity on stratospheric and tropospheric parameters using different complexity of atmospheric models (see Haigh, 2003 for the review) showed that near the latitude of about 50N there is a change of the sign of the tropospheric and stratospheric response to the solar forcing. Another reason for expecting a difference in the response of the portuguese and north-european climates is the effect that the ocean has on the climate in Portugal. Also, the changes in location of the Azores High during the year and the influence of the tropical wind and oceanic currents are important for explaining specific features of the portuguese climate (Luterbacher and Xoplaki, 2003; Miranda et al., 2002; Sáenz et al., 2001a, 2001b; Sánchez et al., 2007). The atmosphere-ocean coupling (e.g. through the changing of the ocean heating regime and, consequently, the wind direction) probably appears in the dependence of the temperature of the current month on the NAOI of the previous month found both by Sánchez et al. (2007) and by us in this study.

A more detailed analysis of the climate in Portugal, especially the analysis of the local stratospheric parameters, could help to separate the influence of the CR and the UV radiation on the troposphere and stratosphere.

## 7. Conclusion

The variations of the temperature in Portugal (Lisbon and Coimbra), the NAOI, the sunspot numbers and the cosmic rays flux over a 140 years period (from 1865 to 2005) have been studied. Different statistical methods (the correlation analysis and the multiple regression analysis) were used. These different methods give similar results and confirm the reliability of each other. The results presented in this paper show that the temperature measured in Portugal depends not only on the internal atmospheric oscillations in the North-Atlantic region but also correlates with the variations of the solar parameters.

The correlations between the temperature in this region and the NAOI depend on the month under consideration and, probably, could be explained by such features of the Portuguese climate as strong influence of the different modes of the circulations (EA and NAO) and the variability of the neighbouring Azores High. These correlations are stronger during the cold months (October to March) and some of the warm months (June to September). For some months there are strong correlations between the temperature variations of the current month and the NAOI values of the previous month (January and May). This effect could be associated with, for example, the atmosphere-ocean coupling. Moderately significant correlations between the NAOI and the sunspot numbers were found only during the cold season.

The correlation between the climatic and the solar parameters show up best on the decadal and decadal-to-centennial timescale (smoothed data). The correlations are stronger during the cold season (November to February). However, the significance of such correlation coefficients, as a rule, is not high (in most cases, p-level > 0.1) due to taking into account the reduction of the number of independent measurements because of the smoothing procedure. On the year-to-year time scale these correlations disappear under the internal atmospheric fluctuations. The predominance of the effect of the decadal-to-centennial solar activity variations on the portuguese climate could be associated with some regional climate features: e.g. the conflicting influence of the different circulation patterns (EA and NAO).

The sign of the correlations between the temperature and solar parameters does not depend on the season: there is the positive correlation with the sunspot numbers and there is the negative correlation with the CR flux intensity throughout the year. The absolute values of the correlation coefficients between the temperature and the CR are higher than those between the temperature and the sunspots (with a small number of exceptions). Besides the CR appears to be the most influential parameter for the multiple regression analysis of the temperature data. The possible explanation is that the sunspots (and the sunspot numbers, consequently) show the main solar activity level and could be used mainly as a proxies for the real agents (e.g. CR, TSI or UV and X-ray irradiance) whereas particles of the CR could be a real active agent, directly affecting atmospheric parameters (levels of ozone, aerosols, cloud amount etc.).

The results of this study could be explained using one (or both) of the following mechanisms of the solar activity influence on the Earth's weather and climate: 1. through the effect of the UV radiations on the stratosphere and stratosphere-troposphere coupling; 2. through the effect of the CR particles on the cloudiness and stratospheric and tropospheric conditions. More data (especially on the local stratospheric parameters) are necessary to conclude which of the mechanisms is more influential.

**Acknowledgement**

We are indebted to Dr. Ivo Alves and M. J. Chorro, from the Geophysical Institute of the University of Coimbra (IGUC), for kindly providing monthly values of temperature from 1865 to 2005 and to Dr. Maria Antónia Valente, from Geophysical Institute Infante D. Luis in Lisbon for kindly providing monthly averaged temperature values from 1855 to 2008. We thank Dr. João Manuel Fernandes for all the support and encouragement he provided and for useful comments that helped to improve the manuscript.
We also acknowledge the University of New Hampshire, 'National Science Foundation Grant ATM-0339527' for making available the Climax Neutron Monitor.

**References**

Abdi, H., 2007. Bonferroni and Šidák corrections for multiple comparisons, in: Salkind, N.J. (Eds.), Encyclopedia of Measurement and Statistics. Thousand Oaks, CA: Sage.
Alanko-Huotari, K., Usoskin, I.G., Mursula, K., Kovaltsov, G.A., 2006. Global heliospheric parameters and cosmic ray modulation: an empirical relation for the last decades. Sol. Phys. 238, 391-404.
Baldwin, M.P., Dunkerton, T.J., 2005. The solar cycle and stratosphere troposphere dynamical coupling. JASTP. 67, 1-2, 71-82.
Bershadskii, A, 2009. Transitional solar dynamics and global warming. Physica A. 388, 15-16, 3213-3224.
Blackford J.J., Chambers, F.M., 1995. Proxy climate record for the last 1000 years from Irish blanket peat and a possible link to solar variability, Earth and Planet. Sci. Let. 133, 145-150
Carslaw, K.S., Harrison, R.G., Kirkby, J., 2002 Cosmic Rays, Clouds, and Climate. Sci. 298, 1732-1737.
Dobrica, V., Demetrescu, C., Boroneant, C., Maris, G., 2009. Solar and geomagnetic activity effects on climate at regional and global scales: Case study - Romania. JASTP. 71, 17-18, 1727-1735.
Fröhlich, C., Lean, J., 1998a. Total Solar Irradiance Variations, in: Deubner, F.L. et al., (Eds.), New Eyes to see inside the Sun and Stars, Proc. IAU Symp. 185, Kyoto, August 1997, Kluwer Acad. Publ., Dordrecht, The Netherlands, pp. 89-102.
Fröhlich, C., Lean, J., 1998b. The Suns Total Irradiance: Cycles, Trends and Related Climate Change Uncertainties since 1978. Geophys. Res. Let. 25, 4377-4380.
Fröhlich, C., 2000. Observations of Irradiance Variations. Sp. Sci. Rev. 94, 15-24
Fröhlich, C., 2003. Long-Term Behavior of Space Radiometers. Metrol. 40, 1, S60-S65
Fröhlich, C., 2004. Solar Irradiance Variability. Geophys. Monogr. Ser. **ISSN** 0065-8448, 141, 97-110
Fröhlich, C., Spruit, H., Wigley, T.M.L., 2006. Variations in solar luminosity and their effect on the Earth's climate. Nat. 443, 7108, 161-166.

Gil-Alana, L.A., 2009. Time Series Modeling of Sunspot Numbers Using Long-Range Cyclical Dependence. Sol. Phys. 257, 2, 371-381
Gleisner, H., Thejll, P., 2003. Patterns of tropospheric response to solar variability. GRL, 30, 13, 44-1, CiteID 1711, DOI 10.1029/2003GL017129.
Haigh, J.D., 2003. The effects of solar variability on the Earth's climate. Phil. Trans. R. Soc. Lond. A, 361, 95-111.
Haigh, J., 2009. Mechanisms for solar influence on the Earth's climate, in: Tsuda, T., Fujii, R., Shibata, K., Geller, M.A. (Eds.), Climate and Weather of the Sun-Earth System (CAWSES), , pp. 231-256.
Hood, L.L., 2004. Effects of solar UV variability on the stratosphere, in: Pap, J. et al. (Eds.) Solar variability and its effect on the Earth's atmosphere and climate system. AGU Monogr. Ser., Washington D.C., pp. 283-303
Hsieh, W.W., 2008. Machine Learning Methods in the Environmental Sciences: Neural Networks and Kernels, Cambridge Univ. Press.
Hurrell, J.W., Kushnir, Y., Ottersen, G., Visbeck M. (Eds.), 2003. The North Atlantic Oscillation: Climatic Significance and Environmental Impact. Geophys. Monogr. Ser., 134.
Holton, J.R., Haynes, P.H., McIntyre, M.E., Douglass, A.R., Rood, R.B., Pfister, L., 1995. Stratosphere-troposphere exchange. Rev. Geophys., 33, 403-439.
Kniveton, D.R., 2004. Precipitation, cloud cover and Forbush decreases in galactic cosmic rays. JASTP, 66, 3-14, 1135-1142.
Kodera, K., Kuroda, Y., 2002. Dynamical response to the solar cycle. J. Geophys. Res. 107(D24), 4749, doi: 4710.1029/2002JD002224.
Krivolutsky, A.A., 2003. History of cosmic ray influence on ozone layer-key steps. Adv. Sp. Res. 31, 9, 2127-2138.
Krivolutsky, A.A., Kuminov, A.A., Repnev, A.I., 1999. Influence of Cosmic Rays on the Earth's Ozonosphere. Geomagn. Aeron. 39, 3, 271-282.
Kristjánsson, J.E., Stjern, C.W., Stordal, F., Fjæraa, A.M., Myhre, G., Jónasson, K., 2008. Cosmic rays, cloud condensation nuclei and clouds – a reassessment using MODIS data, Chem. Phys. 8, 7373–7387
Labitzke, K., 1987. Sunspots, the QBO, and the stratospheric temperature in the north polar region. GRL, 14, 5, 535-537.
Labitzke, K., 2005. On the solar cycle QBO relationship: a summary. JASTP, 67, 1-2, 45-54.
Labitzke, K., van Loon, H., 1988. Associations between the 11-year solar cycle, the QBO and the atmosphere. I - The troposphere and stratosphere in the Northern Hemisphere in winter. JATP, 50, 197-206.
Lawrence, E.N., 1965. Terrestrial climate and the solar cycle. Weather, 20, 334-343.
Lean, J., 1997. The Sun's variable radiation and its relevance for Earth. Annu. Rev. Astron. Astrophys., 35, 33-67.
Lean, J.L., 2010. Cycles and trends in solar irradiance and climate. Wiley Interdiscip. Rev.: Climate Change, 1, 1, 111-122.
Le Mouël, J.-L., Blanter, E., Shnirman, M., Courtillot, V., 2009. Evidence for solar forcing in variability of temperatures and pressures in Europe. JASTP, 71, 12, 1309-1321.
Li, J., Wang, J., 2003. A new North Atlantic Oscillation index and its variability. Adv. Atmos. Sci., 20(5), 661-676.
Luterbacher, J., Xoplaki, E., 2003. 500-Year Winter Temperature and Precipitation Variability over the Mediterranean Area and its Connection to the Large-Scale Atmospheric Circulation, in: Bolle, H.-J. (Ed.) Mediterranean Climate. Variability and Trends, Springer Verlag, Berlin, Heidelberg, pp. 133-153.
Marsh, N., Svensmark, H., 2000. Cosmic Rays, Clouds, and Climate. Sp. Sci. Rev. 94, 1/2, 215-223.
Marsh, N., Svensmark, H., 2003. Solar Influence on Earth's Climate. Sp. Sci. Rev. 107, 1, 317-325.

Miranda, P., Coelho, F.E.S., Tom, A.R., Valente, M.A., 2002. 20th century portuguese climate and climate scenarios, in: Santos, F.D., Forbes, K., Moita, R. (Eds.) Climate change in Portugal: scenarios, impacts and adaptation measures. Gradiva, Lisboa, pp 23–83.
Morozova, A.L., Pudovkin, M.I., Thejll, P., 2002. Variations of atmospheric pressure during solar proton events and Forbush decreases for different latitudinal and synoptic zones. Int. J. Geomagn. Aeron. 3, 2, 181-189.
Naujokat, B., 1986. An update of the observed quasi-biennial oscillation of the stratospheric winds over the tropics. J. Atmos. Sci., 43, 1873-1877.
Ney, E.P., 1959. Cosmic Radiation and the Weather. Nat. 183, 4659, 451.
Pudovkin, M.I., Veretenenko, S.V., Pellinen, R., Kyrö, E., 1996. Cosmic ray variation effects in the temperature of the high-latitude atmosphere. Adv. Space Res. 17, 11, 165-168.
Pudovkin, M.I., Veretenenko, S.V., Pellinen, R., Kyrö, E., 1997. Meteorological characteristic changes in the high-latitude atmosphere associated with Forbush-decreases of the galactic cosmic rays. Adv. Space Res. 20, 6, 1169-1172.
Prieto, L., Garcia, R., Diaz, J., Hernandez, E., del Teso, T., 2002. NAO influence on extreme winter temperatures in Madrid (Spain). Ann. Geophys. 20, 12, 2077-2085.
Rind, D., 2002. The Sun's Role in Climate Variations. Sci. 296, 5568, 673-678.
Sánchez, R.F., Relvas, P., Delgado, M., 2007. Coupled ocean wind and sea surface temperature patterns off the western Iberian Peninsula. J. Mar. Syst. 68, 1-2, 103-127.
Sáenz, J., Rodríguez-Puebla, C., Fernández, J., Zubillaga, J., 2001a. Interpretation of interannual winter temperature variations over southwestern Europe. JGR, 106, D18, 20641-20652.
Sáenz J., Zubillaga, J., Rodríguez-Puebla, C., 2001b. Interannual winter temperature variability in the north of the Iberian Peninsula. Clim. Res. 16, 169-179.
Smith, R.C., Stammerjohn, S.E., Baker, K.S., 1996. Surface air temperature variations in the western Antarctic Peninsula region. Antarct. Res. Ser., 70, 105-121.
Sloan, T., Wolfendale, A.W., 2008. Testing the proposed causal link between cosmic rays and cloud cover. Environ. Res. Let. 3, 2, 024001
Slonosky, V., Yiou, P., 2002. Does the NAOI represent zonal flow? The influence of the NAO on North Atlantic surface temperature. Clim. Dyn. 19, 1, 17-30.
Svensmark, H., Bondo, T., Svensmark, J., 2009. Cosmic ray decreases affect atmospheric aerosols and clouds. GRL, 36, 15, CiteID L1510.
Svensmark, H., Friis-Christensen, E., 1997. Variation of cosmic ray flux and global cloud coverage-a missing link in solar-climate relationships. JATP, 59, 1225-1232.
Tinsley, B.A., Deen, G.W., 1991. Apparent tropospheric response to Mev-Gev particle flux variations: a connection via electrofreezing of supercooled water in high-level clouds? JGR, 96, D12, 22,283-22,296.
Theiler, J., Prichard, D., 1996. Constrained-Realization Monte-Carlo Method for Hypothesis Testing. Physica D 94, 221-235.
Thejll, P., Christiansen, B., Gleisner, H., 2003. On correlations between the North Atlantic Oscillation, geopotential heights, and geomagnetic activity. GRL, 30, 6, 80-1, CiteID 1347, doi10.1029/2002GL016598
Trigo, I.F., Davies, T.D., Bigg, G.R., 1999. Objective Climatology of Cyclones in the Mediterranean Region. J. Clim. 12, 6, 1685-1696.
Tsiropoula, G., 2003. Signatures of solar activity variability in meteorological parameters. JASTP, 65, 4, 469-482.
Usoskin, I.G., Korte, M., Kovaltsov, G.A., 2008. Role of centennial geomagnetic changes in local atmospheric ionization. Geophys. Res. Lett. 35, L05811, doi:10.1029/2007GL033040.
Usoskin, I.G., Alanko-Huotari, K., Kovaltsov, G.A., Mursula, K., 2005. Heliospheric modulation of cosmic rays: Monthly reconstruction for 1951-2004. J. Geophys. Res., 110(A12), A12108.1-A12108.10.

Usoskin, I.G., Mursula, K., Solanki, S.K., Schuessler, M., Kovaltsov, G.A., 2002. A physical reconstruction of cosmic ray intensity since 1610. J. Geophys. Res., 107(A11), 1374.
Veretenenko, S., Thejll, P., 2004. Effects of energetic solar proton events on the cyclone development in the North Atlantic. JASTP, 66, 5, 393-405.
Visbeck, M.H., Hurrell, J.W., Polvani, L., Cullen H.M., 2001. The North Atlantic Oscillation: Past, present, and future. Proc. Natl. Acad. Sci. 98, 23, 12876-12877.
Voiculescu, M., Usoskin, I.G., Mursula, K., 2006. Different response of clouds to solar input. GRL, 33, L21802.1-L21802.5
von Storch, H., 1995. Misuses of statistical analysis in climate research, in: von Storch, H., Navarra, A. (Eds.) Analysis of Climate Variability: Applications of Statistical Techniques, Springer-Verlag, New York, pp. 11-26.
Walsh, K.J.E., Kleeman, R., 1997. Predicting decadal variations in Atlantic tropical cyclone numbers and Australian rainfall, GRL, 24, 24, 3249-3252.
Xoplaki, E., 2002. Climate variability over the Mediterranean. PhD thesis, University of Bern, Switzerland, 195 pp., available at: http://www.giub.unibe.ch/klimet/docs/phd_xoplaki.pdf

**Figures:**

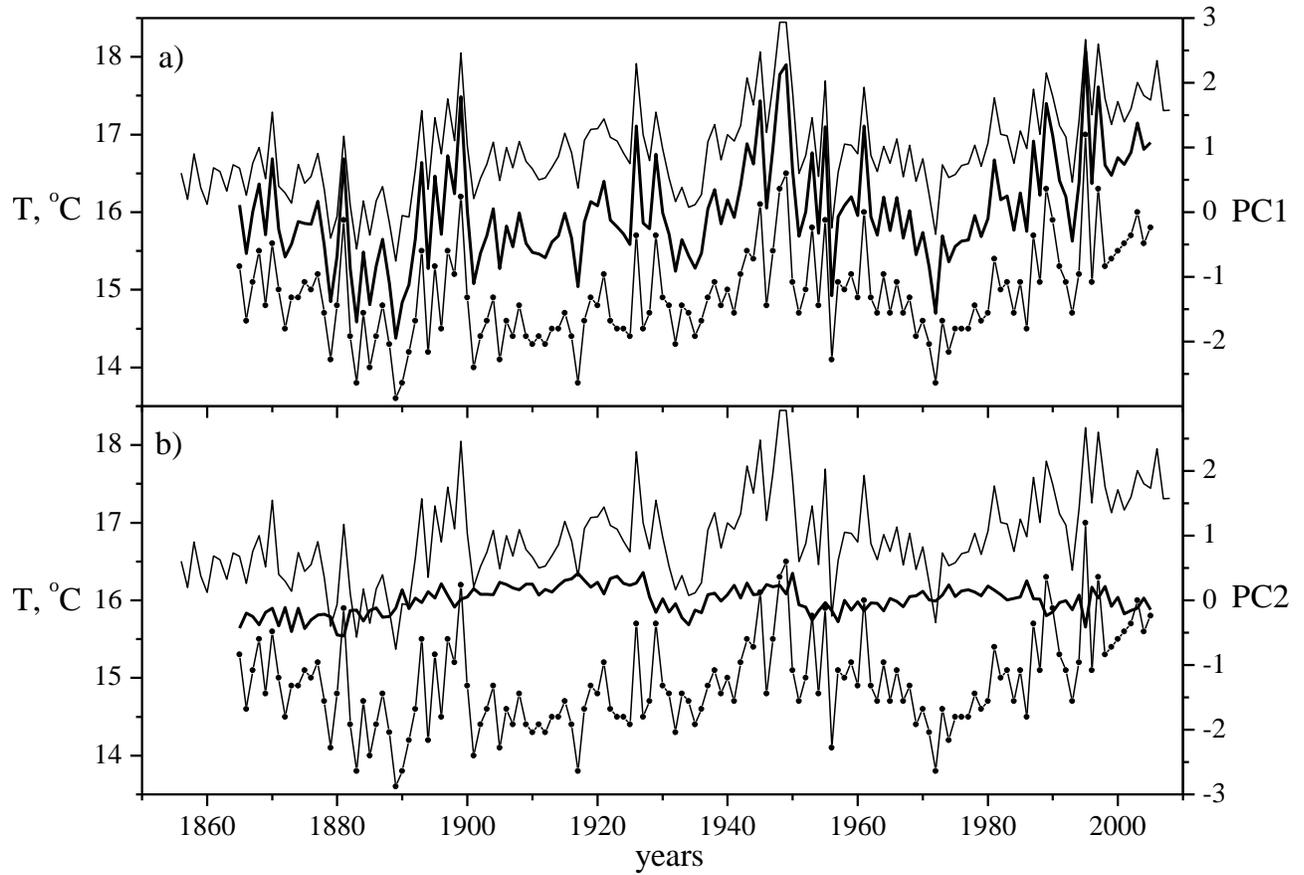

**Figure 1.** Annual temperature variations in Coimbra (lines with dots) and Lisbon (thin lines) and their principal components: a) PC1 and b) PC2 (thick lines).

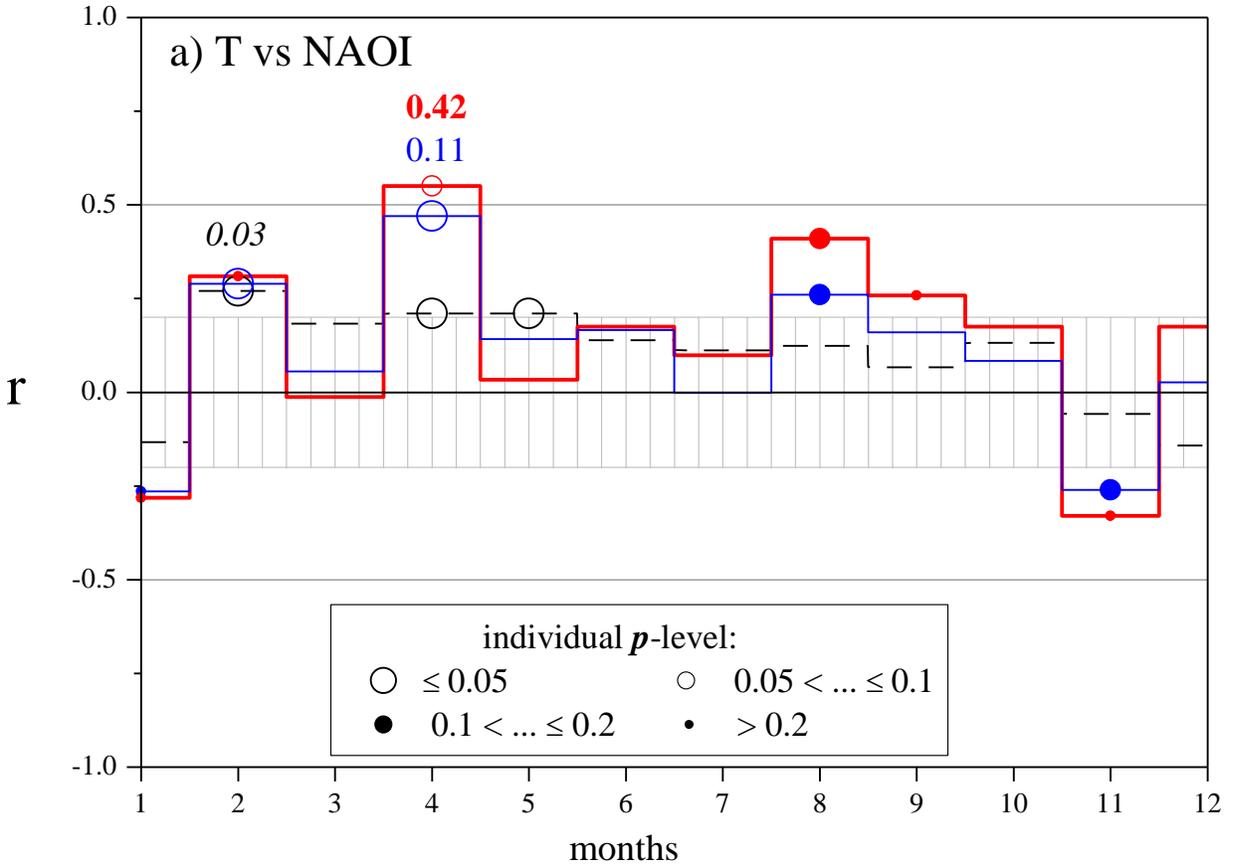

**Figure 2a**. Correlation coefficients (*r*) between the variations of the temperature and the NAOI: non-smoothed data (dashed black line) and smoothed by 5 (thin blue line) and 11 years (thick red line) adjacent averaging. The individual significances (*p*-level) of the $|r| \geq 0.2$ are shown by the circles of different size and fills. The area $|r| \leq 0.2$ is shaded. Numbers show the batch significances of the largest correlation coefficients for each of three data series: non-smoothed data (in black italic) and smoothed by 5 (in blue) and 11 years (in bold red).

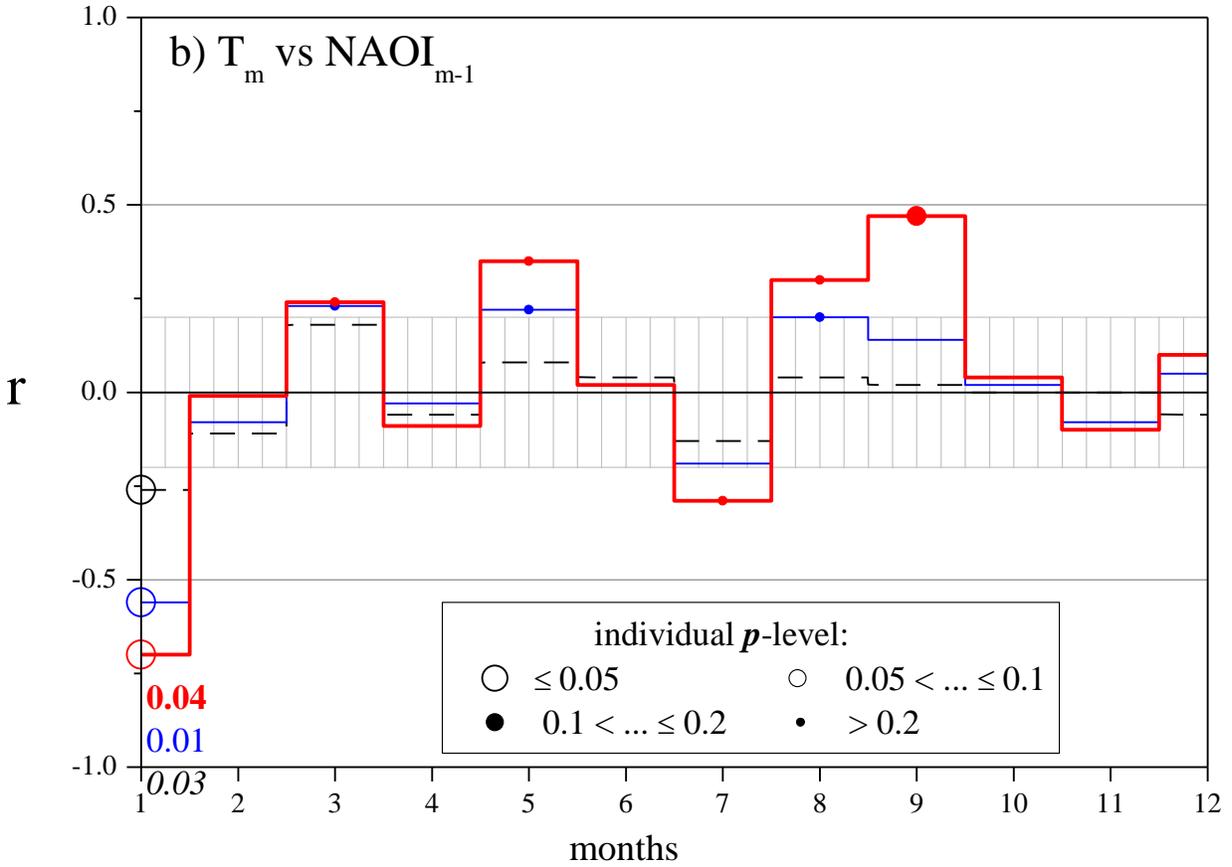

**Figure 2b**. Correlation coefficients (*r*) between the variations of the temperature of the current month ($T_m$) and the NAOI of the previous month ($NAOI_{m-1}$): non-smoothed data (dashed black line) and smoothed by 5 (thin blue line) and 11 years (thick red line) adjacent averaging. The individual significances (p-level) of the $|r| \geq 0.2$ are shown by the circles of different size and fills. The area $|r| \leq 0.2$ is shaded. Numbers show the batch significances of the largest correlation coefficients for each of three data series: non-smoothed data (in black italic) and smoothed by 5 (in blue) and 11 years (in bold red).

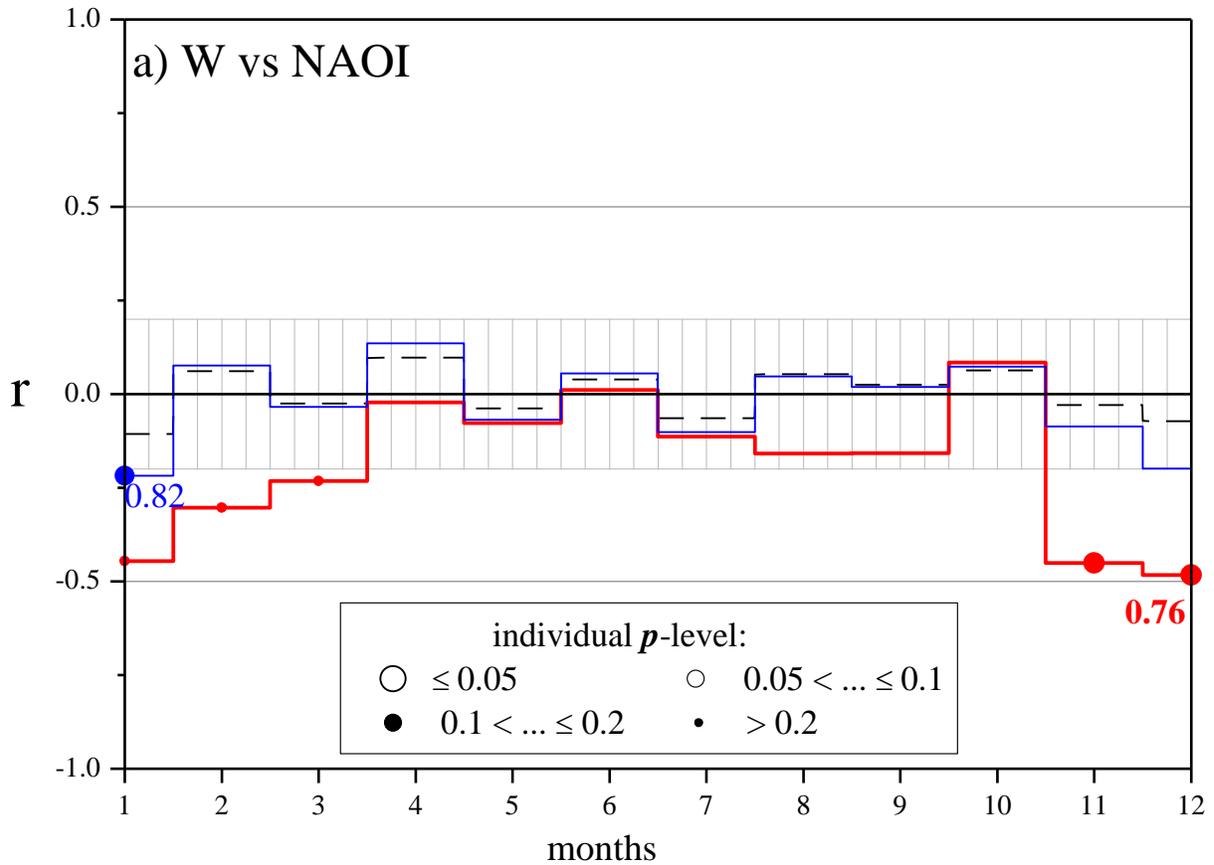

**Figure 3a**. Correlation coefficients (*r*) between the variations of the NAOI and the sunspot numbers: non-smoothed data (dashed black line) and smoothed by 5 (thin blue line) and 11 years (thick red line) adjacent averaging. The individual significances (*p*-level) of the $|r| \geq 0.2$ are shown by the circles of different size and fills. The area $|r| \leq 0.2$ is shaded. Numbers show the batch significances of the largest correlation coefficients for each of three data series: non-smoothed data (in black italic) and smoothed by 5 (in blue) and 11 years (in bold red).

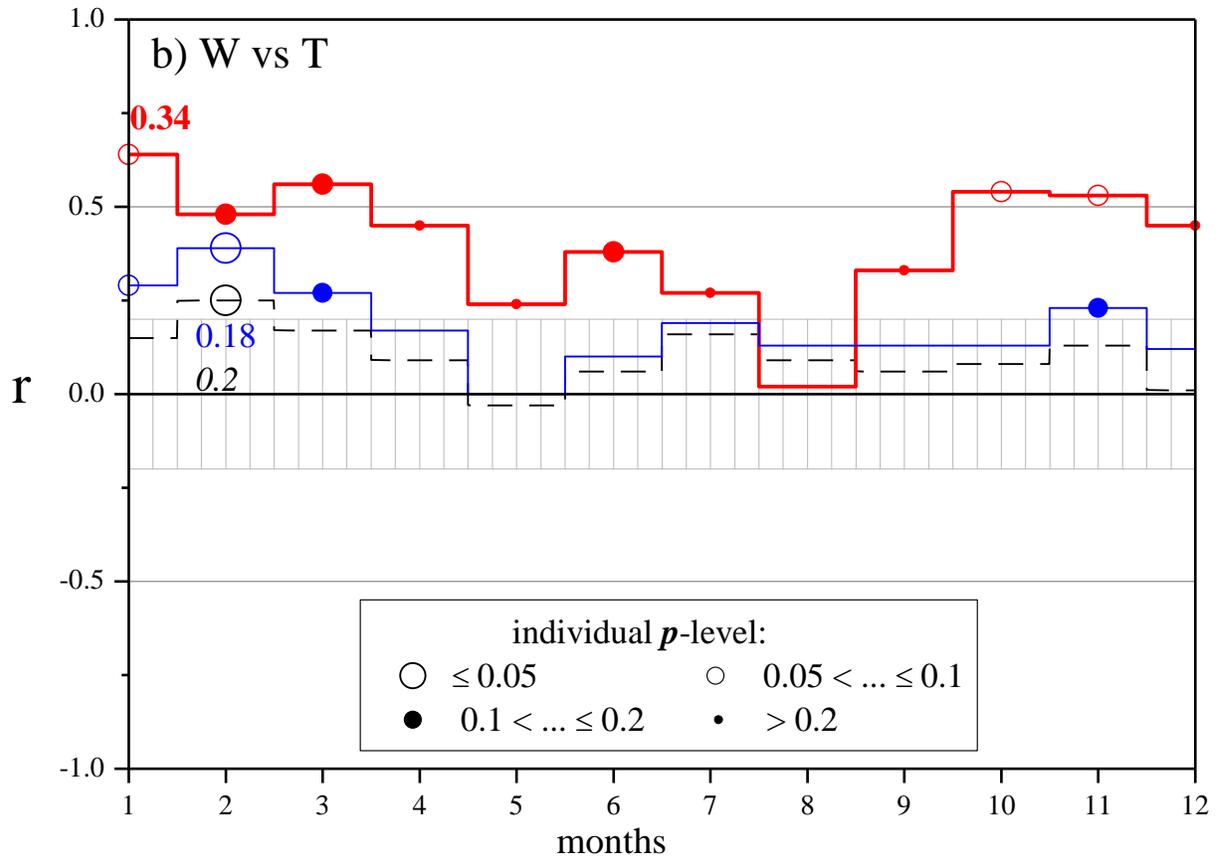

**Figure 3b**. Correlation coefficients (*r*) between the variations of the temperature and the sunspot numbers: non-smoothed data (dashed black line) and smoothed by 5 (thin blue line) and 11 years (thick red line) adjacent averaging. The individual significances (*p*-level) of the $|r| \geq 0.2$ are shown by the circles of different size and fills. The area $|r| \leq 0.2$ is shaded. Numbers show the batch significances of the largest correlation coefficients for each of three data series: non-smoothed data (in black italic) and smoothed by 5 (in blue) and 11 years (in bold red).

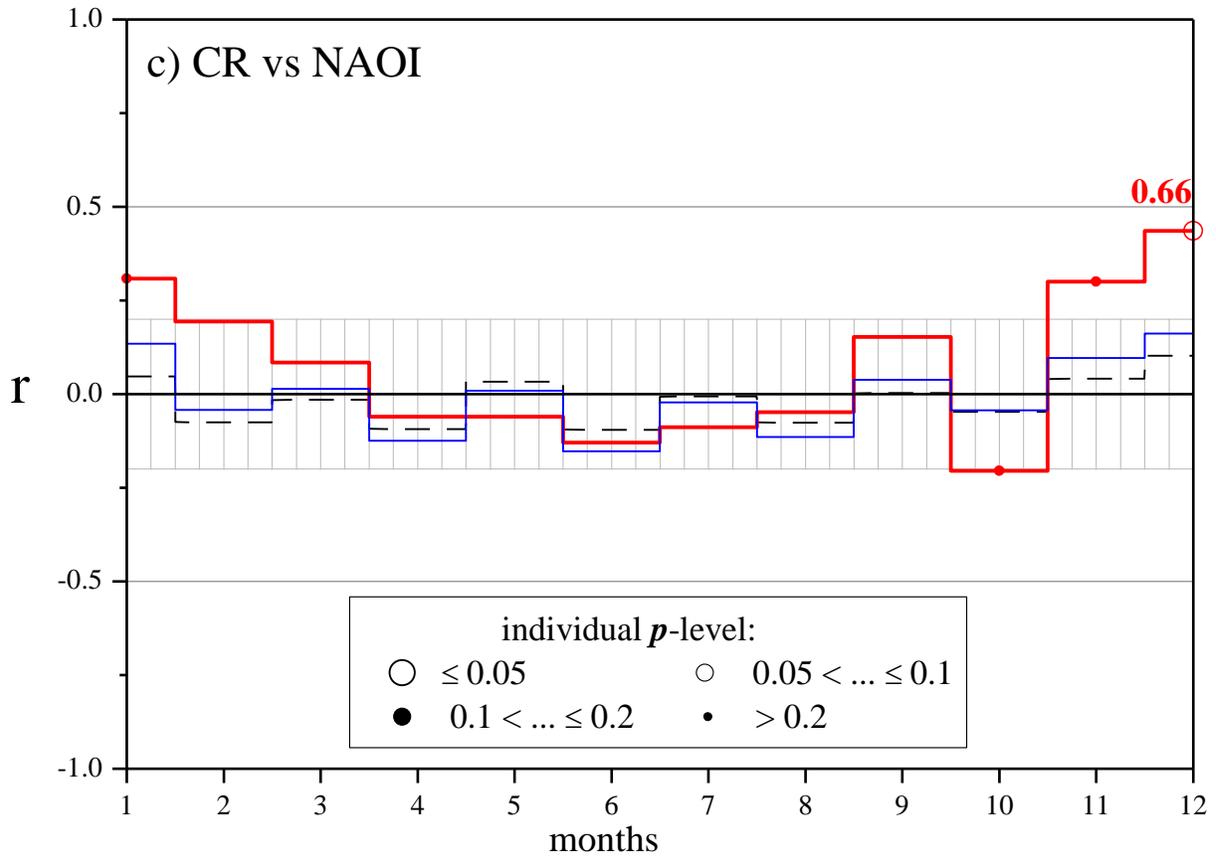

**Figure 3c**. Correlation coefficients (*r*) between the variations of the NAOI and CR: non-smoothed data (dashed black line) and smoothed by 5 (thin blue line) and 11 years (thick red line) adjacent averaging. The individual significances (*p*-level) of the $|r| \geq 0.2$ are shown by the circles of different size and fills. The area $|r| \leq 0.2$ is shaded. Numbers show the batch significances of the largest correlation coefficients for each of three data series: non-smoothed data (in black italic) and smoothed by 5 (in blue) and 11 years (in bold red).

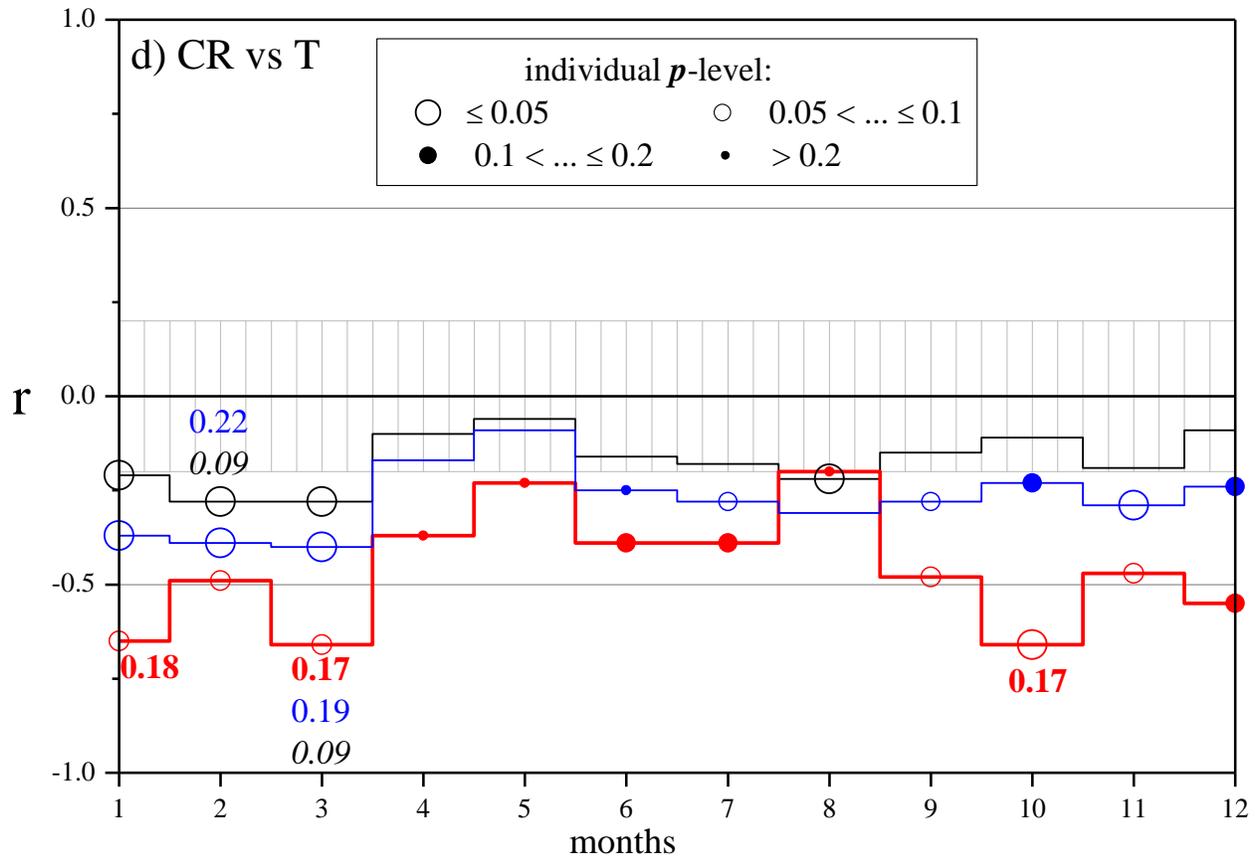

**Figure 3d.** Correlation coefficients (*r*) between the variations of the temperature and CR: non-smoothed data (dashed black line) and smoothed by 5 (thin blue line) and 11 years (thick red line) adjacent averaging. The individual significances (*p*-level) of the $|r| \geq 0.2$ are shown by the circles of different size and fills. The area $|r| \leq 0.2$ is shaded. Numbers show the batch significances of the largest correlation coefficients for each of three data series: non-smoothed data (in black italic) and smoothed by 5 (in blue) and 11 years (in bold red).

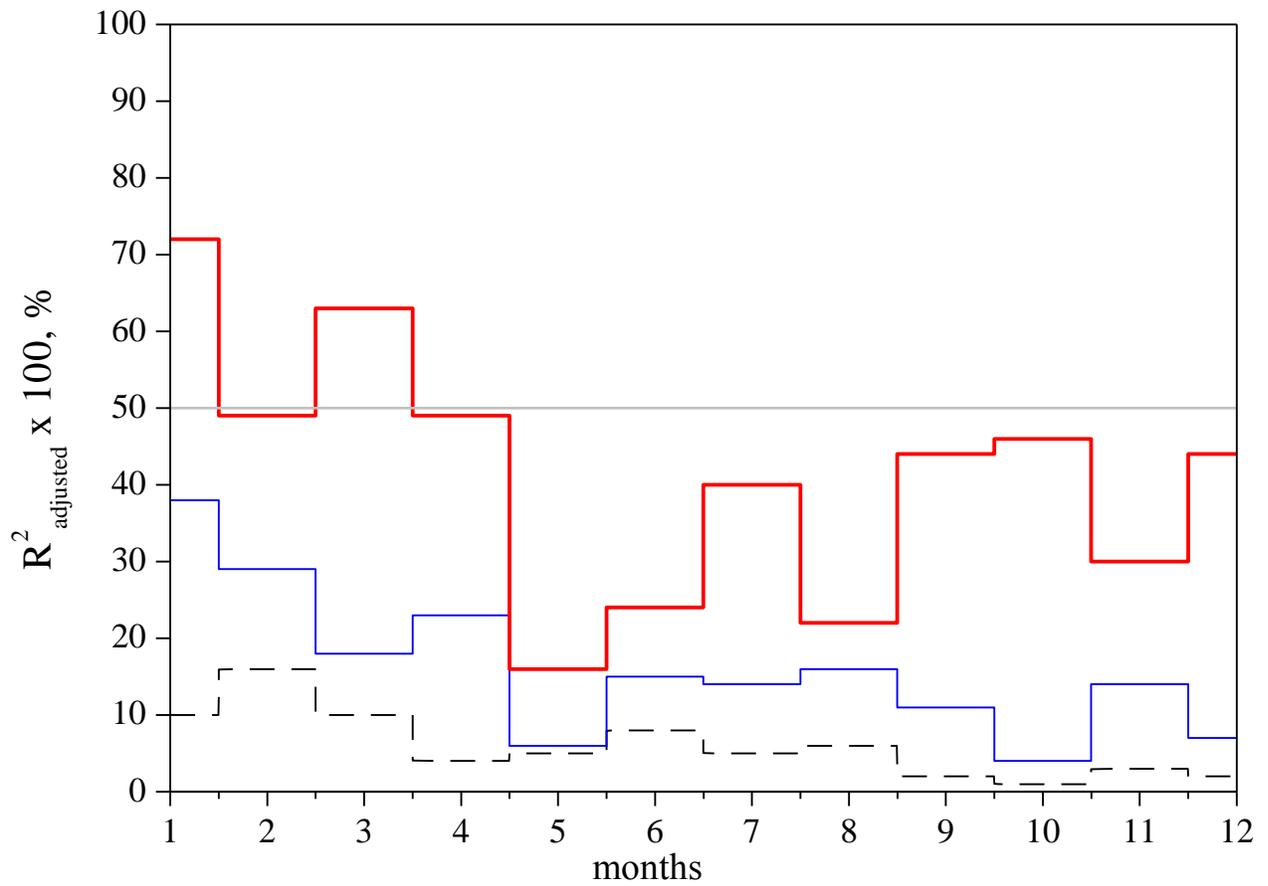

**Figure 4**. ($R^2_{adjusted} \times 100$) values - percent of the variations of the temperature accounted for by the multiple regression model for the particular month: non-smoothed data (black dashed line), and smoothed by 5 (thin blue line) and 11 years (thick red line) adjacent averaging.

**Tables:**

**Table 1.** Correlation coefficients between the sunspot numbers and both measured ($CR_{Climax\ NM}$) and reconstructed ($CR_{reconstruction}$) cosmic ray flux; $p$-level < 0.01.

|  | $CR_{Climax\ NM}$ | $CR_{reconstruction}$ |
|---|---|---|
| W | -0.73 ÷ -0.82 | -0.8 |
| $CR_{Climax\ NM}$ | – | > 0.99 |

**Table 2.** Correlation coefficients between the atmospheric and the solar parameters for "cold" and "warm" seasons; individual significances ($p$-levels) are in brackets. Only $p$-level for $|r| \geq 0.2$ are shown, $|r| \geq 0.5$ are in bold.

| season | W vs T | CR vs T | T vs NAOI | W vs NAOI | CR vs NAOI |
|---|---|---|---|---|---|
| "cold" | 0.28 (0.009) | -0.36 (<0.001) | 0.0 | 0.0 | 0.0 |
| "warm" | 0.13 | -0.3 (<0.001) | 0.11 | 0.0 | 0.0 |
| "cold", 5 yrs. adj.av. | 0.42 (0.009) | **-0.5** (<0.001) | 0.0 | -0.14 | 0.1 |
| "warm", 5 yrs. adj.av. | 0.22 (0.2) | -0.41 (0.007) | 0.28 (0.24) | 0.0 | 0.0 |
| "cold", 11 yrs. adj.av. | **0.67** (0.05) | **-0.7** (0.02) | 0.0 | **-0.53** (0.14) | 0.4 (0.13) |
| "warm", 11 yrs. adj.av. | 0.28 (0.46) | -0.44 (0.15) | 0.36 (0.26) | -0.17 | 0.0 |

**Table 3.** Parameter sets defined by the "best subset" technique which are used in the multiple regression analysis. "+" indicate the parameters which are actually included in each specific regression model. The analysis is performed for both non-smoothed and smoothed data for different months and seasons. Shaded cells indicate the parameters excluded from specific regression models.

| month | non-smoothed data | | | | 5 points adjacent averaged | | | | 11 points adjacent averaged | | | |
|---|---|---|---|---|---|---|---|---|---|---|---|---|
| | $NAOI_m$ | $NAOI_{m-1}$ | CR | W | $NAOI_m$ | $NAOI_{m-1}$ | CR | W | $NAOI_m$ | $NAOI_{m-1}$ | CR | W |
| 1 | | + | + | | | + | + | | + | + | + | |
| 2 | + | + | + | | + | + | + | | + | + | + | + |
| 3 | + | + | + | | | + | + | | + | + | + | |
| 4 | + | | + | | + | + | + | | + | | + | + |
| 5 | + | | + | + | + | + | + | + | | + | | + |
| 6 | + | | + | + | + | | + | + | + | + | + | + |
| 7 | + | + | + | | + | + | + | + | + | + | + | + |
| 8 | + | | + | + | + | | + | + | + | + | + | + |
| 9 | | | + | + | + | | + | + | + | + | + | + |
| 10 | | | + | + | | | + | | | | + | + |
| 11 | | | + | | + | | + | | | | + | + |
| 12 | + | | + | | | + | + | | + | + | + | |
| "cold" season | | ▨ | + | | | ▨ | + | | + | ▨ | + | + |
| "warm" season | | ▨ | + | + | + | ▨ | + | + | + | ▨ | + | + |

**Table 4.** ($R^2_{adjusted} \times 100$) values - percent of the variability of the temperature accounted for by the multiple regression model for the particular season.

| season | $R^2$adjusted × 100, % |
|---|---|
| "cold" | 12 |
| "warm" | 11 |
| "cold", 5 yrs. adj.av. | 25 |
| "warm", 5 yrs. adj.av. | 24 |
| "cold", 11 yrs. adj.av. | 60 |
| "warm", 11 yrs. adj.av. | 37 |